\documentclass[12pt]{article}
\usepackage{epsf}
\usepackage{epsfig}
\usepackage{graphicx}

\begin{document}

%%% THIS IS FOR CDF NOTE ONLY 
\begin{titlepage}
%\begin{flushright}
%CDF/PUB/EXOTIC/PUBLIC/4238\\
%\end{flushright}
\vspace*{3cm}
%%%%%%%

\begin{center} 
{\bf SEARCH FOR NEW PARTICLES} \\ 
{\bf IN MULTIJET FINAL STATES AT THE TEVATRON}\\
\mbox{} \\
\mbox{} \\
Tommaso~Dorigo\\
for the CDF and D0 Collaborations \\
{\em Padova University, Via Marzolo 8, I-35131 PADOVA, Italy}  \\
\mbox{} \\
\mbox{} \\
\end{center}
      
\begin{center}
\textbf{Abstract}
\end{center}
We present the latest results of the searches for new particles in hadronic
final states performed in $p\bar p$ collisions at $\sqrt{s}=1.8$~TeV. \par
The large data samples collected with the CDF and D0 detectors at the Tevatron
collider between 1992 and 1995 allow searches for low cross section 
phenomena in dijet and multijet events, despite the hindrance of the 
high background from normal QCD processes. However, no signal for new physics
is found, and the data show good agreement with QCD. Limits on 
the mass and on the cross section of the searched states can thus be set. 
\end{titlepage}

\vspace {1cm}

\section{Introduction}

Ever since the discovery of the W and Z bosons with the UA1 and UA2 detectors in
1983, proton--antiproton colliders have proven to be successful 
probes for new physics in the high energy domain. The high rates and high
collision energies achievable in these environments have pushed our 
understanding of the strong interactions and of particle physics in 
general to farther and farther horizons. The usefulness of hadronic
probes in particle interactions has however always come together with an
enormous complexity of the final states, as compared with the clean,
easily understandable and manageable physics electron-positron
colliders can provide. But leptons are not only a guarantee for a clean
interaction when they collide: they have always been thoroughly exploited as a
unmistakable signal for high interest phenomena when they show up in
the detectors. $p\bar p$ colliders currently use high $P_T$ 
leptons as the best way to trigger and to tag particle decays as
W and Z bosons, heavy quarks like charm, bottom and top, and to search
for supersymmetry and new phenomena.

New physics searches in the zero lepton final states are a hard thing to
manage. One may just compare the production cross section of a pair of
jets with 30 GeV of transverse energy --which is in the microbarns
ballpark at the Tevatron-- with the same final state caused by the decay
of a W or Z boson, which lies in the few nanobarns domain: this example
may give the idea of the hard times one usually faces when trying to use
these final states for new particle searches.    

\begin {figure} [h]
\begin {minipage}{0.49\linewidth}
\centerline{\epsfig{file=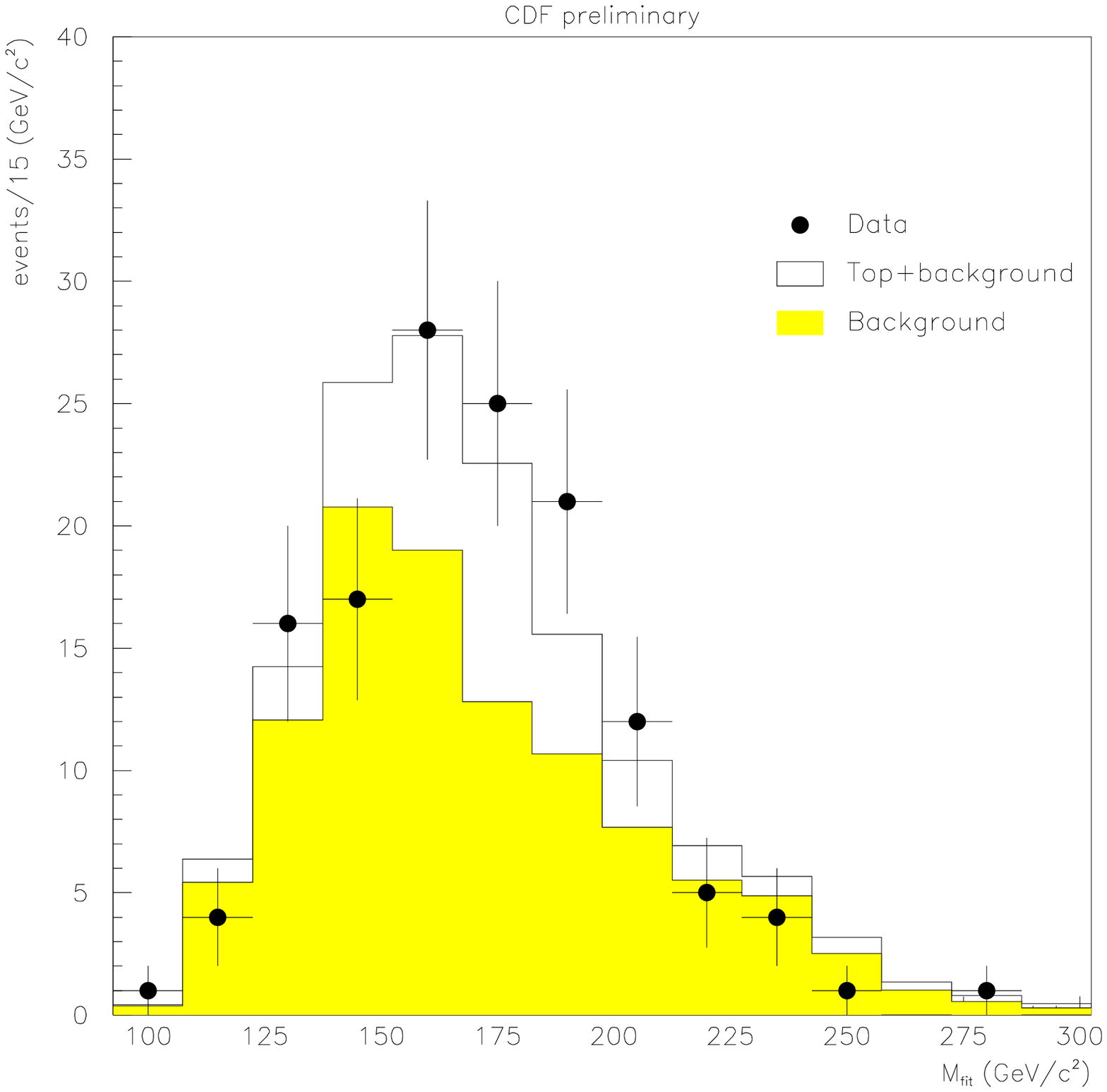,
bb=0 0 600 600,width=8cm,clip=}} 
\caption{ {\em Reconstructed mass of the top quark from events with six jets.
          The black dots are data, the shaded histogram is the expected 
          background
          contribution, and the white histogram is the sum of background and}
          $t\bar{t}$ {\em contributions.} }
\label {figure:tt6j}
\end{minipage}
\hspace {0.2cm}
\begin {minipage}{0.49\linewidth}
\centerline{\epsfig{file=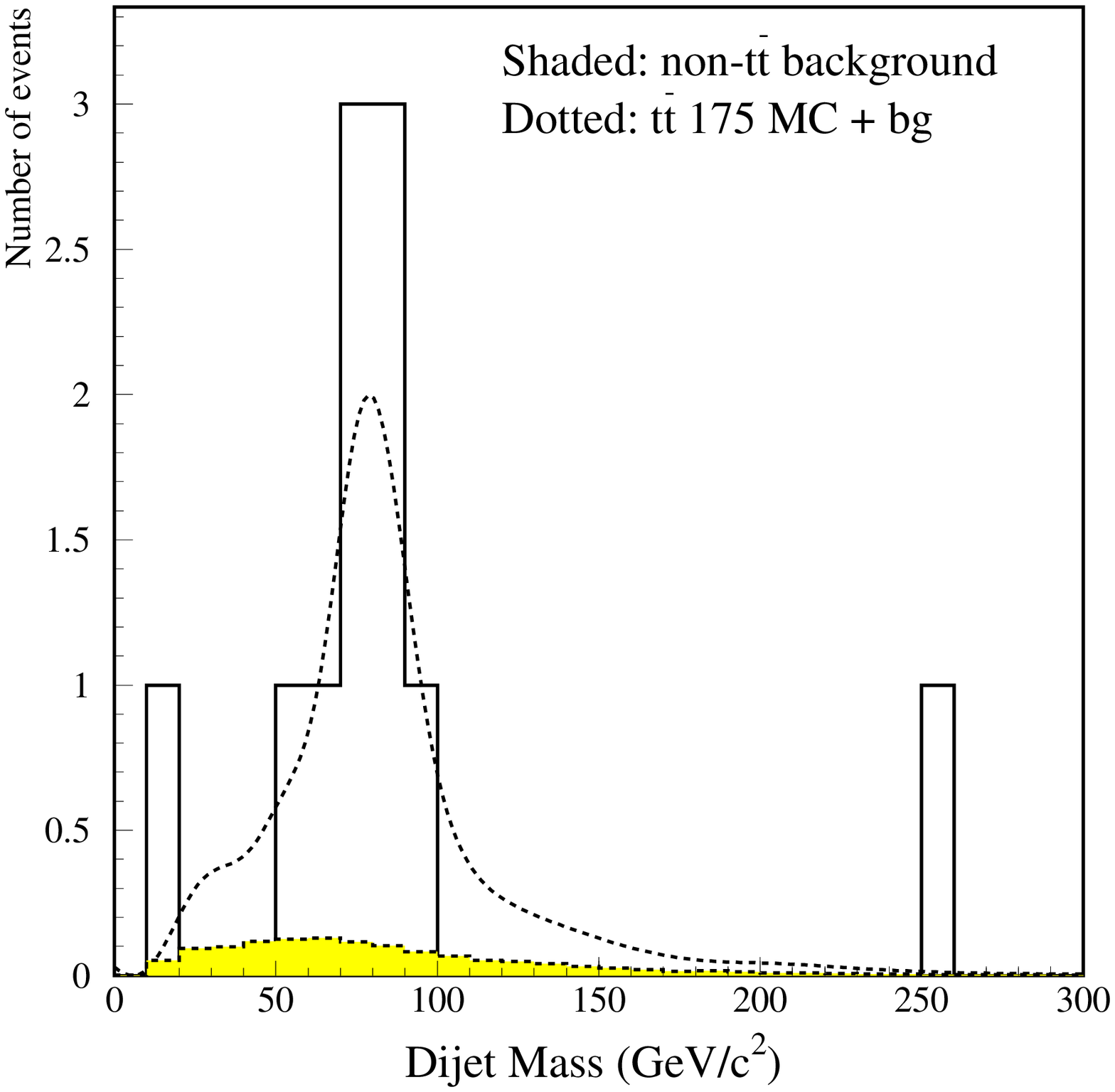,
bb=0 150 600 750,width=8cm,clip=}} 
\caption{ {\em The dijet mass distribution for events with an electron or muon,
          missing transverse energy, and two more jets both coming from b 
          decay. The 11 events are compared to background expectations 
          alone (shaded) and to background plus} $t\bar{t}$.}
\label {figure:Wjj}
\end{minipage}
\end {figure}

Notwithstanding the low signal/noise ratio, a vector boson signal 
has been published in 1987 by the UA2 collaboration\footnote
  {{\em Phys. Lett.} 186B (1987), 452. The search
  was later updated with a larger data sample which produced a signal of $5367\pm958$ events:
  {\em Z. Phys.} C49 (1991), 17.}.
For many years since then no other resonances in jet final
states have shown up in 
hadronically triggered data samples at $p\bar p$ colliders. But things
are rapidly changing now: the CDF collaboration has used a multijet data
sample to observe the six jet decays of the $t\bar t$ pairs 
(see fig.\ref{figure:tt6j});
and another CDF analysis to be submitted to {\em Physics Review Letters}
has recently shown how a clean $W\rightarrow jj$ peak can be extracted from
a sample enriched in $t\bar t$ decays (see fig.\ref{figure:Wjj}).

The next collider run at the Tevatron, predicted to start in 1999,
will hopefully exploit the advantageous branching ratios of   
hadronic decays of many new states to push still further their
searches: the next
millennium will probably start with Tevatron and LHC showing how jets can
be very effective in reconstructing the event 
decay kinematics when TeV physics comes into play.

\section {Jets with the CDF and D0 Detectors}

   The CDF and D0 collaborations have put a considerable effort in    
designing and building detectors that could measure with sufficient
precision jet energies in the largest possible solid angle, 
being aware of the importance
of these measurements for both QCD studies and for the reconstruction
of high mass decays like those of top quark pairs. \par
Both detectors come equipped
with calorimeters, divided in inner  
electromagnetic sections and larger outer portions designed for 
complete containment of hadronic showers; and both are physically divided into
sections that cover different ranges of pseudorapidity, up to $|\eta|\simeq4$. 
CDF features a central scintillator-based calorimeter and 
forward-backward symmetric devices where a Argon-Ethane gas mixture is the active medium; 
the segmentation geometry in $\eta-\phi$ space is $0.1\times(0.09 \div 0.27)$, 
and the
stochastic resolution term is $\sigma_E/E=14\%/\sqrt{E}$ for electrons, and
$\sigma_E/E=75\%/\sqrt{E} $ for hadrons.

   D0 features a uniform design of depleted Uranium and liquid Argon calorimetry
both for the central region and for the two endcap detectors. 
The calorimeter segmentation is $0.1 \times 0.1$ in most of
the detector components, but is finer in the third layer of the 
electromagnetic sampling section ($0.05 \times 0.05$), the one where e.m. showers
approach their maximum development. For electrons the resolution is 
$\sigma_E/E=15\%/\sqrt{E} $, and for hadrons is about 
$\sigma_E/E=50\%/\sqrt{E} $.
\par
To trigger on jet events, both experiments rely on many single cluster
triggers, hardwired with different energy thresholds. 
The lower energy triggers have to be heavily prescaled in order to keep 
the rate of event collection matched to data storage capabilities. 
Both collaborations have also designed multijet triggers able to collect 
events with four or more jets, and triggers based on the sum of transverse
energy in the entire calorimeter. 
\par
To reconstruct jets offline, both D0 and CDF use cone-based algorithms that 
fulfil the Snowmass conference standards\footnote{
  J.Huth et Al. in "Research Directions for the decade"
  Proceedings of the summer study on High energy Physics,
  Snowmass, Colorado, 1990, ed. E.L.Berger (World Scientific,1991).}.
The jet radius is normally chosen to be R=0.7, a value considered 
to be the best compromise 
between the reduction of underlying event background and the minimization of
out-of-cone losses, and the most stable with regards to theoretical
calculations.

The algorithms devised for jet identification compile lists of towers over a 1 GeV
threshold in transverse energy, and then form circles around them, evaluate the
center of gravity of the energy distribution inside the circles, 
and displace them to those positions;
the procedure is repeated until stability is achieved. 
Jet merging is decided if two neighboring jets
share more than a certain fraction of the lower energy jet (50 \% at D0, 75\% at CDF).
\par
The jet energies have to be carefully corrected before being useful for 
mass bump searches.  
CDF uses a detector response function to take care 
of nonlinearities and of possible energy losses at detector boundaries;
after that, an absolute energy scale correction is evaluated with the help
of Monte Carlo simulations, to take
into account out-of-cone losses, and to correct for underlying event
contributions to the energy flow inside the cones, low response
of the calorimeter to hadrons, and other effects;
a systematic error on the jet energy scale can then be evaluated by 
using Z+1 jet data, where the Z is well measured through its 
leptonic decay and has to balance in $P_T$ the jet.
The D0 collaboration first sets the absolute jet energy scale in 
the e.m. compartments, with the aid of Z decay electrons, and then 
uses photon+jet events to get the response function for the jet in the 
hadronic compartments.  After the corrections, both collaborations estimate 
the uncertainty on the jet energy scale at the 5 \% level. 
 
    An important feature of the CDF detector is its
ability to identify b quark decays by reconstructing their decay vertex, 
thanks to a very precise Silicon VerteX detector (SVX)
capable of a $\sim15 \mu m$ impact parameter resolution for charged tracks.
The D0 detector can also tag b quark decays by 
identifying low $P_T$ muons in the jet cones; this method is however limited by 
the branching ratio for semileptonic decays and by the muon identification efficiency.
 
\section { Heavy Neutral Scalar Searches in Four Jet Events }

At the Tevatron, associated production is predicted to be the 
most promising process for a search of neutral scalar particles,
such as the long-hunted Higgs boson
or new states predicted by MSSM or by technicolor models introduced as 
alternative scenarios for electroweak symmetry breaking, such as light 
color singlet technipions\footnote
   { See for instance the paper by Eitchen, Lane and Womersley, 
     HEP-PH/9704455. }.
 
   The most promising signature for these processes consists in a charged 
lepton plus some missing 
transverse energy due to an undetected neutrino --both from W decay-- 
accompanied by a pair of jets originated from the b quarks the new particle has decayed to. 
Both D0 and CDF have searched in this channel.
These searches have been flanked at CDF by a study of four jet events, to look for 
both the W and the H bosons in the dijet final state. 
The high background from QCD processes can be substantially reduced
with the request that two of the four jets in the event come from b decay.

\begin{figure}[h]
\begin {minipage} {0.49\linewidth}
\centerline{\epsfig{file=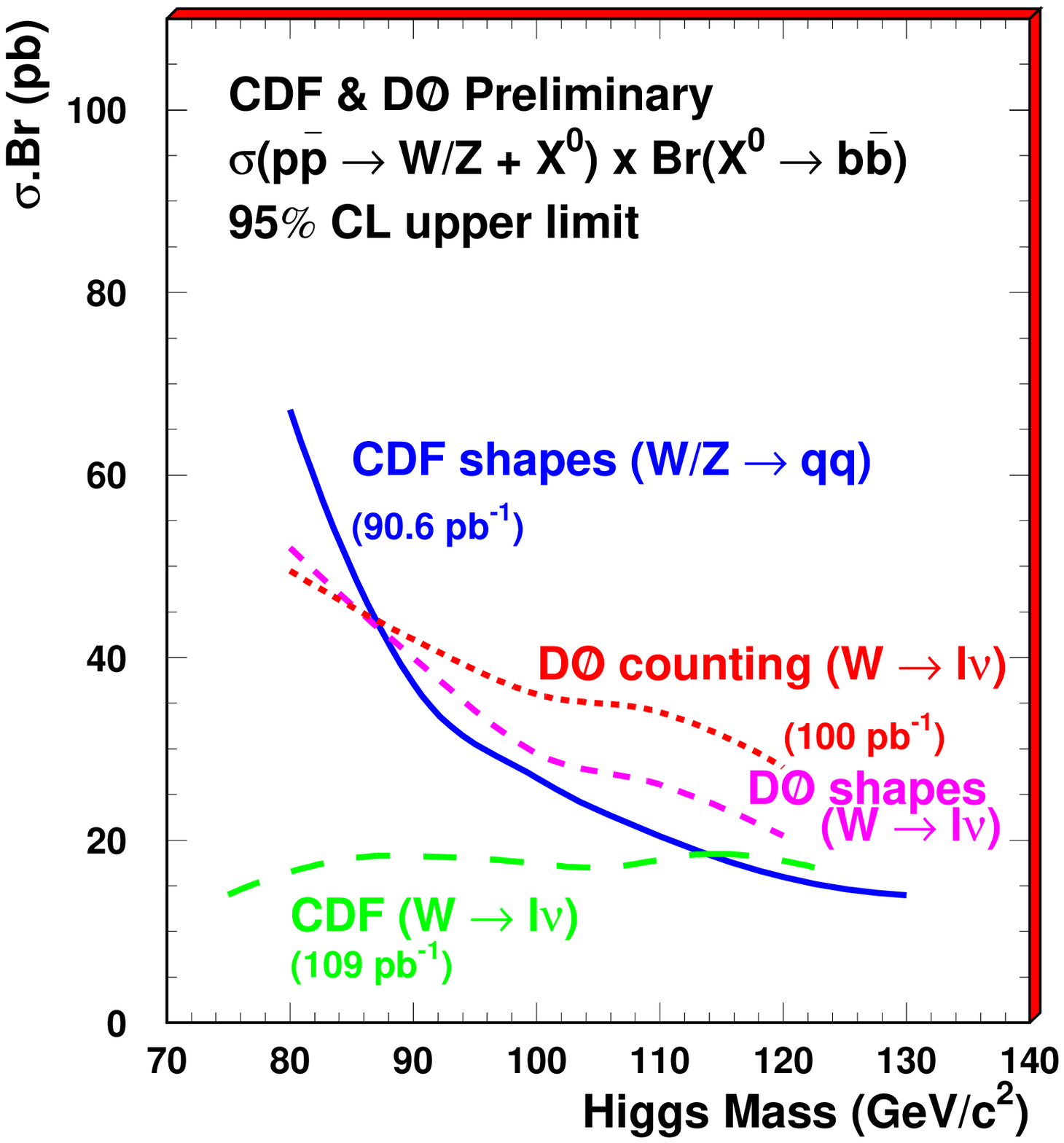,
bb=0 0 500 550,width=8cm,clip=}}
\caption { \em The various cross section limits obtained by D0 and CDF for the
           WH production. }
\label{figure:WHlimit}
\end{minipage}
\hspace{0.2cm}
\begin{minipage}{0.49\linewidth}
\centerline{\epsfig{file=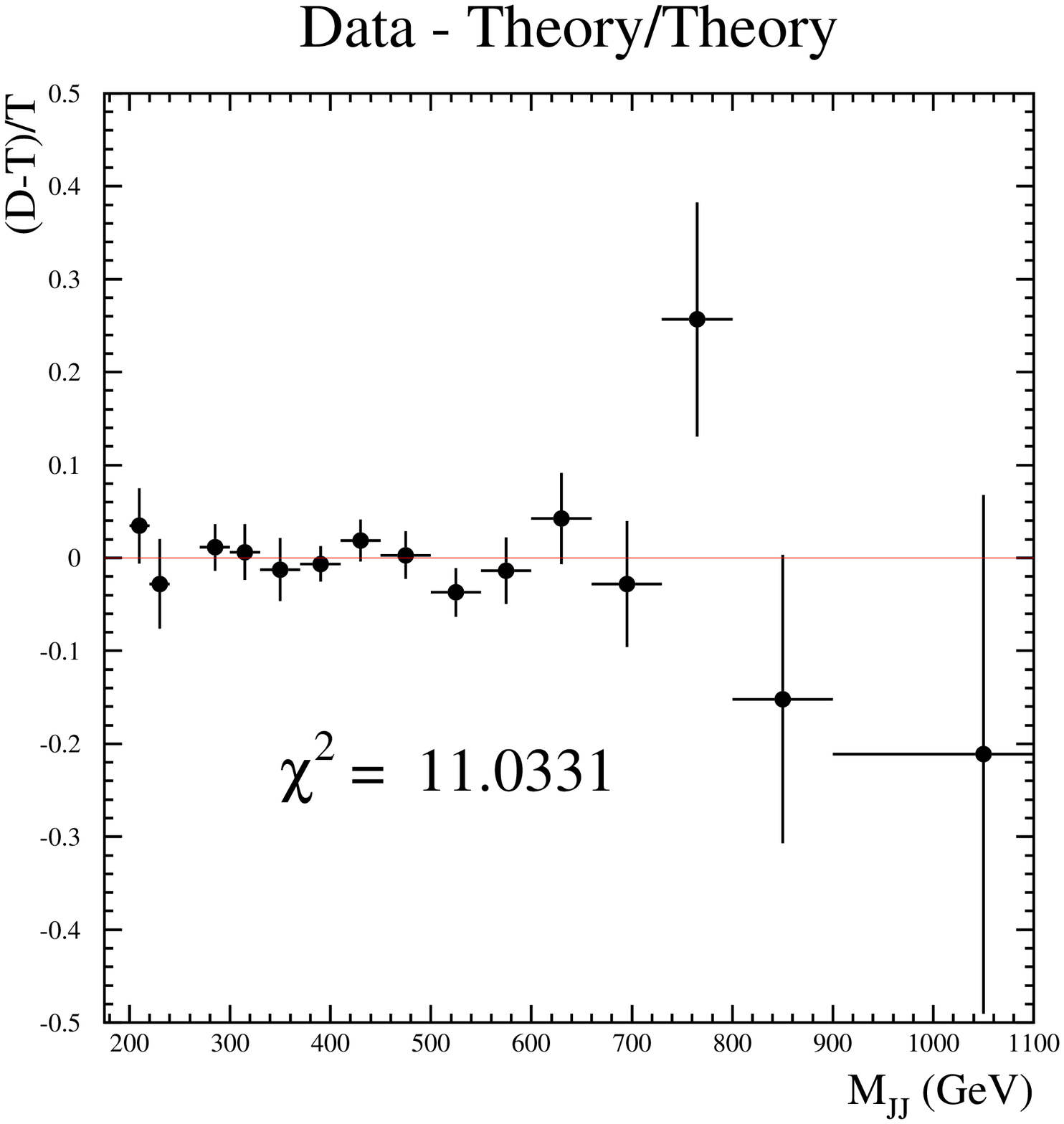,
bb=0 0 600 600,width=8cm,clip=}}
\caption { \em The residuals of the fit of the dijet mass spectrum with the 
           QCD distribution obtained with JETRAD. } 
\label{figure:D0scarti}
\end{minipage}
\end{figure}

   By studying the invariant mass distribution of the b jet pairs, CDF has
put a new limit on the cross section for associated WH production (see
fig.\ref{figure:WHlimit}). For a detailed description of this search the reader
is referred to the writeup of Leslie Groer's talk in these Proceedings.
 
\section {Dijet Bump Searches}

Many extensions to the Standard Model predict the existence of new states
which may decay to a pair of jets. These include axigluons, excited quarks,
color octet technirhos, W' and Z' bosons, and $E_6$ diquarks. 
These states can be searched for in events
with two jets by computing the dijet mass and comparing it with 
background predictions: the absence of bumps
in the spectrum may be turned into cross section limits. 

\subsection {Searches at D0}

To search for excited quarks in the process $qg \rightarrow q^{\ast} \rightarrow qg$, and
for new gauge bosons W' and Z', the D0 collaboration
uses data collected with its four single jet triggers. The data samples, 
featuring $E_T$ thresholds at 30, 50, 85 
and 115 GeV, correspond for the 1994-95 data taking to integrated luminosities of 0.355,
4.56, 51.7 and 90.7 $pb^{-1}$, respectively.

%    A removal of isolated noisy calorimeter cells and accelerator losses
%is performed with quality cuts that discard single tower spikes and jets where 
%the fraction of energy collected in the e.m. modules lies outside the interval 
%$0.05 \div 0.95$: events are kept if the two most energetic jets fulfil these requirements.
%After a removal of cosmic ray background based on a cut on missing transverse energy,
%the residual background contamination is estimated to be less than $2\%$ for jets with
%$E_T$ lower than 500 GeV. Finally, a removal of events with three or more simultaneous
%interactions in the same bunch crossing is performed.

   A removal of isolated noisy calorimeter cells, accelerator losses, cosmic ray background
and pileup events is performed with quality cuts; after the removal the residual background
contamination is less than 2 \% for jets with $E_T$ lower than 500 GeV.
For each event that passes the selection, a dijet invariant mass can be calculated.
The events are weighted by the efficiency of the quality cuts applied.
To enhance the S/N ratio for new particles over the QCD
background, the pseudorapidity
of the first two jets is required to lie in the interval $|\eta|<1.0$, 
and their difference
has to be $|\eta_1-\eta_2|<1.6$. The relative normalizations of the four datasets
are established by requiring equal cross sections in the regions where they overlap. 
The four datasets are used above mass thresholds of 200, 270, 370 and 500
GeV, where they are fully efficient.

    From the mass spectrum a resonance contribution can be estimated by fitting the data
with a QCD continuum distribution obtained with the JETRAD Monte Carlo\footnote 
{W.Giele, E.Glover, D.Kosover, {\em Phys. Rev. Lett.} {\bf 73} (1994), 2019. }
and a smearing procedure that takes into account the jet resolution, plus a signal line shape 
obtained with the PYTHIA Monte Carlo\footnote 
{H.Bengtsson, T.Sjostrand, {\em Comp. Phys. Comm.} {\bf 46} (1987), 43.}
for excited quarks of a given mass; the fit is 
performed with the binned maximum likelihood method. 
The QCD shape alone is able to fit the data well, as can be seen in 
fig.{\ref{figure:D0scarti}. 
By taking into account the systematic uncertainties in the cross section, 
coming from the uncertainties in integrated luminosity ($8\%$), 
jet energy scale ($5\%$) 
and data selection cuts ($2\%$), a cross section limit can be obtained
(fig.\ref{figure:D0crosslimit}).
%    The cross section is obtained with the formula 
%$\sigma_X \times BR \times a = N_X / {\cal{L}}$, where $a$ is the acceptance
%and ${\cal{L}}$
%is the integrated luminosity. The uncertainty of the cross section is given by 
%the quadrature sum of the measurement error $\Delta N_X$, the luminosity uncertainty ($8\%$),
%and the data selection cuts ($2\%$), while the uncertainty in the jet energy scale ($5\%$), 
%that translates in a similar uncertainty in the mass, 
%is taken into account by plotting the $95\%$ C.L. 
%limit at the mass that produces the most conservative limit. The limit
%is then compared to the expected cross sections for the searched states, times their
%branching ratio and acceptance (see fig.\ref{figure:D0crosslimit}). 
\begin{figure}[h!]
\begin {minipage}{0.49\linewidth}
\centerline{\epsfig{file=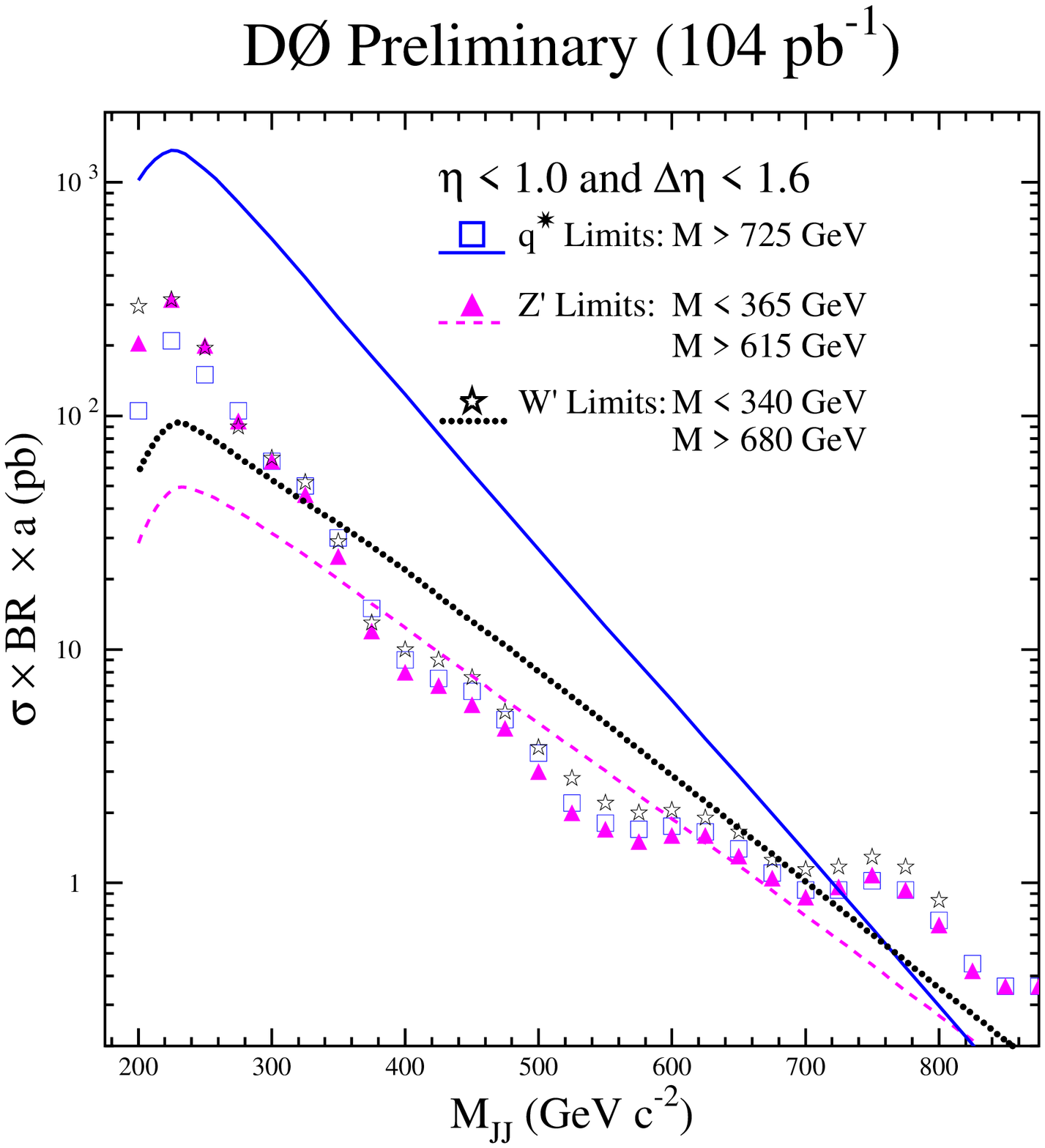,
bb=0 0 600 600,width=8cm,clip=}} 
\caption { \em Mass limits obtained for excited quarks and new gauge bosons
            by the D0 collaboration.}
\label{figure:D0crosslimit}
\end{minipage}
\hspace{0.2cm}
\begin{minipage}{0.49\linewidth}
\centerline{\epsfig{file=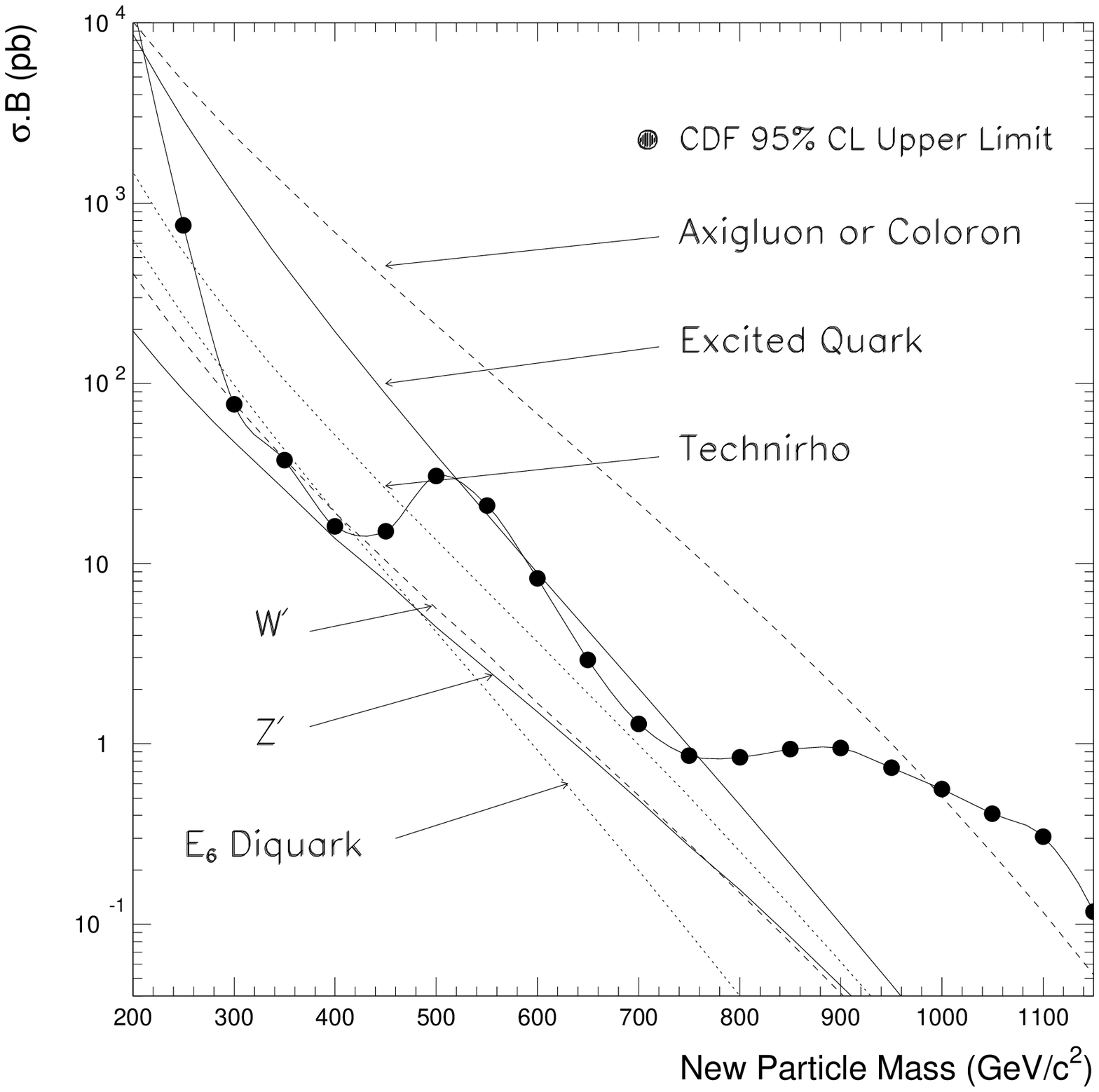,
bb=0 150 600 700,width=8cm,clip=}}
\caption { \em The upper limit on the production cross section times branching 
           ratio of the searched states obtained by the CDF collaboration. }
\label{figure:masslimits}
\end{minipage}
\end{figure}
It is thus possible
to exclude excited quarks with a mass lower than $720$ GeV, 
W' bosons in the range
$340<M_{W'}<680$\ GeV, and Z' bosons in the range $365<M_{Z'}<615$\ GeV. 

\subsection {Searches at CDF}

To search for new states decaying to a pair of jets, CDF uses data collected
with single jet triggers. These have nominal threshold at 20, 50, 70 and 100
GeV, and their effective integrated luminosities are respectively 
0.126, 2.84, 14.1 and 106 $pb^{-1}$.
After jet corrections, the four datasets are used to measure the dijet mass
spectrum above 180, 241, 292 and 388 GeV respectively, that is from the point
where they become fully ($>95\%$) efficient. Events are required to have
the two more energetic jets inside a pseudorapidity region $|\eta|<2.0$, and 
their scattering angle in the center-of-mass frame to be
$|\cos \theta^{\ast}| < 2/3$: this provides uniform acceptance
as a function of mass while reducing the QCD background.
The differential cross section, $d\sigma/dM$, is then plotted  
%(fig.\ref{figure:xscdf}) 
versus the mean
dijet mass in bins of width approximately equal to the dijet mass resolution
($RMS \sim 10\%$), and fit with a parametrization $d\sigma/dm = 
A(1-m/\sqrt{s}+Cm^2/s)^N/m^P$, with four parameters A,C,N,P. This gives an
adequate description of the spectrum ($\chi^2/DOF=1.49$).

In the absence of an excess of events over the fit distribution, which for
a new particle would show up in at least two neighboring bins, an upper limit
for the cross section of the new particles is extracted by performing a binned 
likelihood fit of the data to the background parametrization and the mass
resonance shape\footnote 
  { The resonance shape is the same for all the 
    particles considered, since the natural width of each
    is predicted to be smaller than the experimental resolution.}, 
for 20 values of new particles mass ranging from 200 to 1150
GeV in 50 GeV intervals. By convoluting each of the 20 likelihood distributions
with the corresponding total gaussian systematic uncertainty, 95\%
C.L. upper limits in the cross section are extracted. 
The systematic uncertainties come from many different sources,
the most relevant being the energy scale (5\%); 
their effect is evaluated
by varying the source of uncertainty by $\pm 1\sigma$ and refitting the data.
From the predicted lowest order production cross sections for the searched 
particles we can thus extract exclusion intervals in 
their mass (see fig.\ref{figure:masslimits}). 
The exclusion regions are: for axigluons and flavor universal colorons, between
200 and 980 GeV; for excited quarks, between 80 and 570
GeV and between 580 and 760 GeV\footnote 
{Due to a likely statistical fluctuation of the data in the region around 580 GeV, CDF
cannot exclude that mass region for excited quarks; that interval is however ruled out
by the D0 search.};
for color octet technirhos,
between 260 and 480 GeV; for W' bosons, between 300 and 420 GeV; and 
for $E_6$ diquarks, between 290 and 420 GeV.
%These exclusion regions are
%evaluated for Standard Model couplings ($f=f'=f_s=1$). If smaller couplings
%are considered, the excluded mass intervals vary as shown in 
%fig.\ref{figure:massvscoupling}.

\section {Characterization of Multijet Events}

The CDF collaboration has recently published\footnote
{ F.Abe {\em et al.}, {\em Phys. Rev.} {\bf D54} (1996), 4221. }
 a detailed analysis of events with three, four
and five jets, where a set of variables that completely characterize the 
event kinematics
are used to compare the data to QCD predictions and to a phase space model.
We review that analysis here, since it can actually be thought of 
as a search for new physics
in multijet final states, and we also discuss a soon-to-be-published analysis of six jet 
events, performed with the same tools. 
%These searches show that multijet events
%provide a strong confirmation of QCD and that no new physics is necessary to explain
%the data in these final states. 

\subsection { Comparisons of Three, Four and Five Jet Events to QCD Predictions }

   CDF has studied events with three, four and five jets collected 
from 1992 to 1995 with a trigger requiring the sum of jet
transverse energy to be greater than 
300 GeV. Multijet mass distributions and configuration variables have been used to compare
the data to a
leading order QCD matrix element calculation, a QCD parton shower calculation, and a model
where jets are distributed evenly in the N-body phase space.
  
   The leading order QCD predictions were obtained with the NJETS\footnote
{F.A.Behrends, W.Giele, H.Kuijf, {\em Nucl. Phys.} {\bf B333} (1990), 120.}
Monte Carlo program, based on a complete calculation of the LO $2\rightarrow N$ matrix element.
To avoid singularities the program needs as input the minimum separation between the final state
partons, which was chosen to be $\Delta R=0.9$ to match 
the experimental data; the chosen structure
functions were the KMRSD0, with the renormalization scale set at the average $P_T$
of the outgoing partons. The parton energies were then smeared according to the CDF jet energy
resolution, $\sigma_E=0.1 E$. The parton shower calculations were done with the HERWIG 
Monte Carlo\footnote 
{G.Marchesini, B.Webber, G.Abbiendi, I.Knowles, M.Seymour,
 L.Stanco, {\em Comp. Phys. Comm. } 67 (1992), 465.}
, together with a full
simulation of the CDF detector response. The structure functions used were the CTEQ1M, and 
the renormalization scale was set at the value $Q^2=stu/2(s^2+t^2+u^2)$. The phase space 
model was based on the GENBOD phase space generator\footnote {
{\em CERN Prog. Libr. Man.} 1989.10.03, Rout. W515, 6.503. The generator was provided in 
input with the single jet mass distributions and the multijet mass distributions predicted
with the help of the HERWIG Monte Carlo.}.
%, and was provided in input with
%the single jet mass distributions and the multijet mass distributions predicted by the HERWIG
%Monte Carlo. 
Comparisons between phase space distributions and QCD calculations allow an
understanding of what are the variables most sensitive to QCD effects  
in multijet production.

   The starting data sample, featuring events with jets with pseudorapidity $|\eta|<3.0$, with
corrected transverse energy $E_T>20$ GeV, their sum being $\Sigma E_T>420$ GeV, 
consists in 30245 events.
To completely characterize a N jet event one can devise a set of (4N-4) variables\footnote
   { These variables have been introduced in S.Geer and T.Asakawa, 
     {\em Phys. Rev.} {\bf D53} (1996), 4793. }.
The most relevant N-jet variable is the total mass of the N jets, 
and the remaining variables can be reduced by iteratively 
merging together the two jets with lowest dijet mass, and
considering the event to be a N-1 jet event; this procedure can be stopped 
when three objects
remain, where the eight 3-jet variables are well known and used in the 
literature: the 3-jet
mass $M_{3j}$, the four parameters that specify the relative configuration 
$X_3$, $X_4$, 
$\cos {\theta}_3$ and ${\psi}_3$, and  three variables that 
specify the single jet masses,
$f_1$, $f_2$ and $f_3$. The remaining variables for a 4-jet (four) or a 
5-jet (eight) event 
can then be chosen to be the normalized masses of the two merged jets, 
the energy fraction
of the more energetic of the two merged jets, and the cosine of the 
angle between the plane
containing the resultant jet and the beam and the plane containing the 
two jets alone before their merging took place.
 
%   The multijet mass distributions are shown for data and QCD calculations in 
%fig.\ref{figure:njmass}: both Monte Carlo programs are able to reproduce the spectra well.

Both Monte Carlo programs are seen to be able to reproduce the multijet mass distributions
well.
For the three jet variables, the most dramatic disagreement between phase space 
predictions and data is found on the angular distributions in $\cos {\theta}_3$ and 
${\psi}_3$, while QCD predictions are in reasonable agreement with the data. The Dalitz 
variables $X_3$ and $X_4$ are also well reproduced by both QCD Monte Carlos. The single jet
masses are slightly overestimated by the HERWIG calculations.

   For the four and five jet data the distributions overall show that both QCD calculations
are able to reproduce well both the Dalitz and the angular distributions, with a slight
superiority of the leading order NJET Monte Carlo. The phase space model is on the contrary
unable to reproduce the angular distributions, and fails to describe 
correctly also some regions
of the 4- and 5-jet Dalitz distributions.

   From these comparisons, it is possible to see that $2 \rightarrow 2$ scattering plus 
gluon radiation provides a good first approximation to the full LO QCD matrix element for
events with 3, 4 and 5 jets in the final state. The phase space model disagrees most with
the data in the regions of parameter space where QCD predicts large contributions to initial
and final state radiation. 

\subsection { Comparison of Six Jet Events to QCD Predictions }

An analysis similar to that just described has very recently been completed at CDF by
using events
with six jets in the final state. The methodology of the comparison is completely equivalent
to that just reported for three to five jet events; the differences are the following:
\begin {itemize}
\item a slightly different trigger, requiring $\sum E_T>175$ GeV, 
      has been used to collect the six jet events dealt with here;
\item the NJETS LO calculations are exact only with up to five partons 
      in the final state; for
      N=6, NJETS uses an approximation based on neglecting non-LO color
      contributions\footnote{F.Behrends, W.Giele, H.Kuipf, {\em Phys. Lett.}
      {\bf 232B} (1989), 266.} and by simplifying the helicity computation 
      with the SPHEL approximation\footnote{ For a description see
      F.Behrends, H.Kuijf, {\em Nucl. Phys.} {\bf B353} (1991), 59.};
\item different parametrizations have been chosen for the structure functions:
      for the HERWIG Monte Carlo the CTEQ2L were used, while for the 
      NJETS Monte Carlo the CTEQ3M were used;
\item in order to avoid trigger inefficiency regions of phase space, a
      higher cut is applied on the total mass of the six jets: 
      $M_{6j}>520$ GeV. 
\end{itemize}

The data, consisting in 1282 events, shows a good agreement with the QCD
calculations on all the 20 variables necessary to fully describe the 
final state. Many of the distributions are instead very poorly described
by the phase space model, particularly where poles are contributing from
the QCD multijet matrix element, corresponding to soft and collinear
radiation from the incoming and outgoing partons (see for instance figs.
\ref{figure:psi3_6} and \ref{figure:x4}).

\begin {figure} [h!]
\begin{minipage}{0.49\linewidth}
\centerline{\epsfig{file=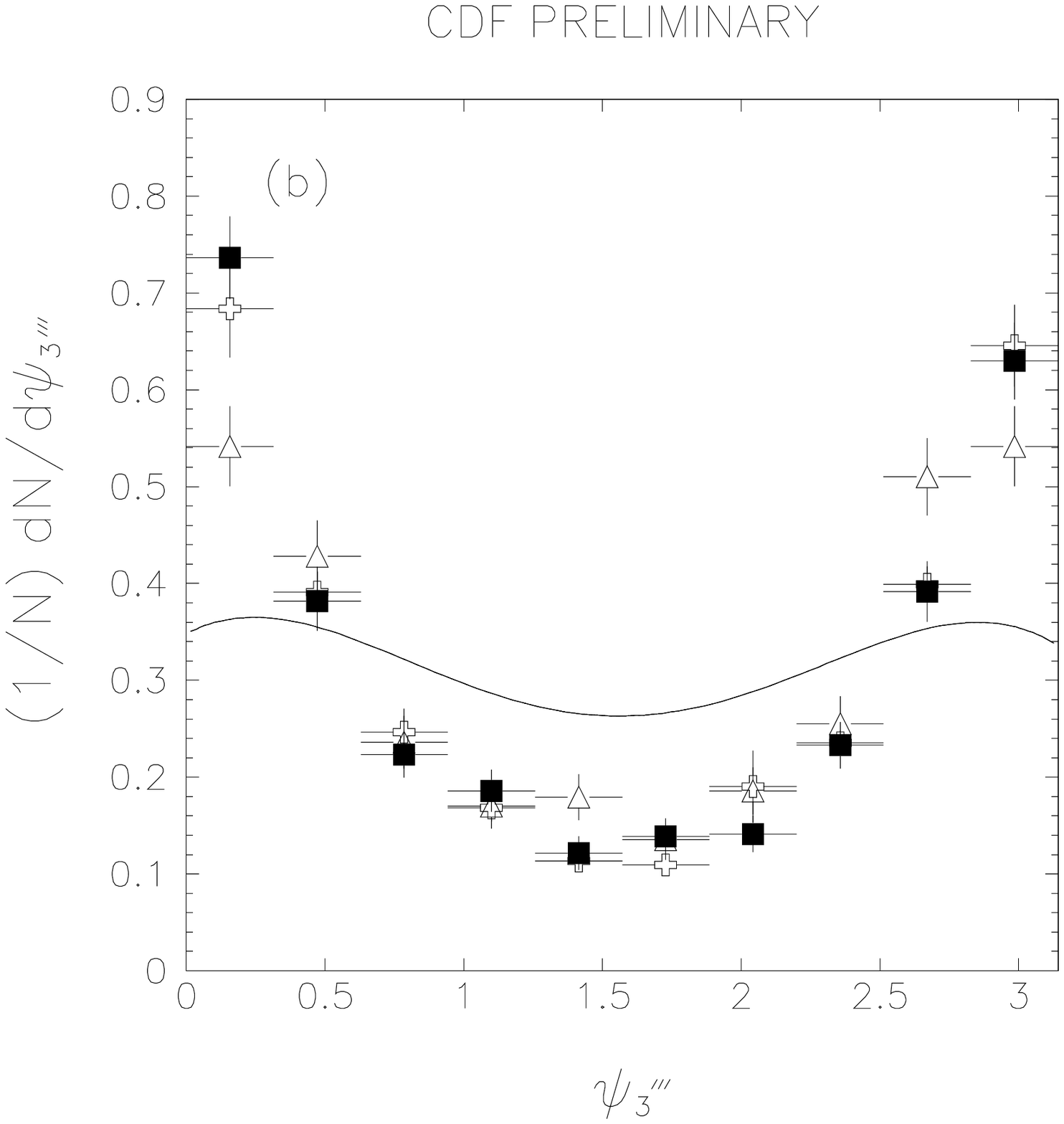,
bb=0 150 550 670,width=8cm,clip=}}
\caption { $\psi_3$ \em distribution for the data (solid squares) compared
           to the predictions of HERWIG (triangles), NJETS (crosses) and
           phase space (smooth line). }
\label{figure:psi3_6}
\end{minipage}
\hspace {0.2cm}
\begin{minipage}{0.49\linewidth}
\centerline{\epsfig{file=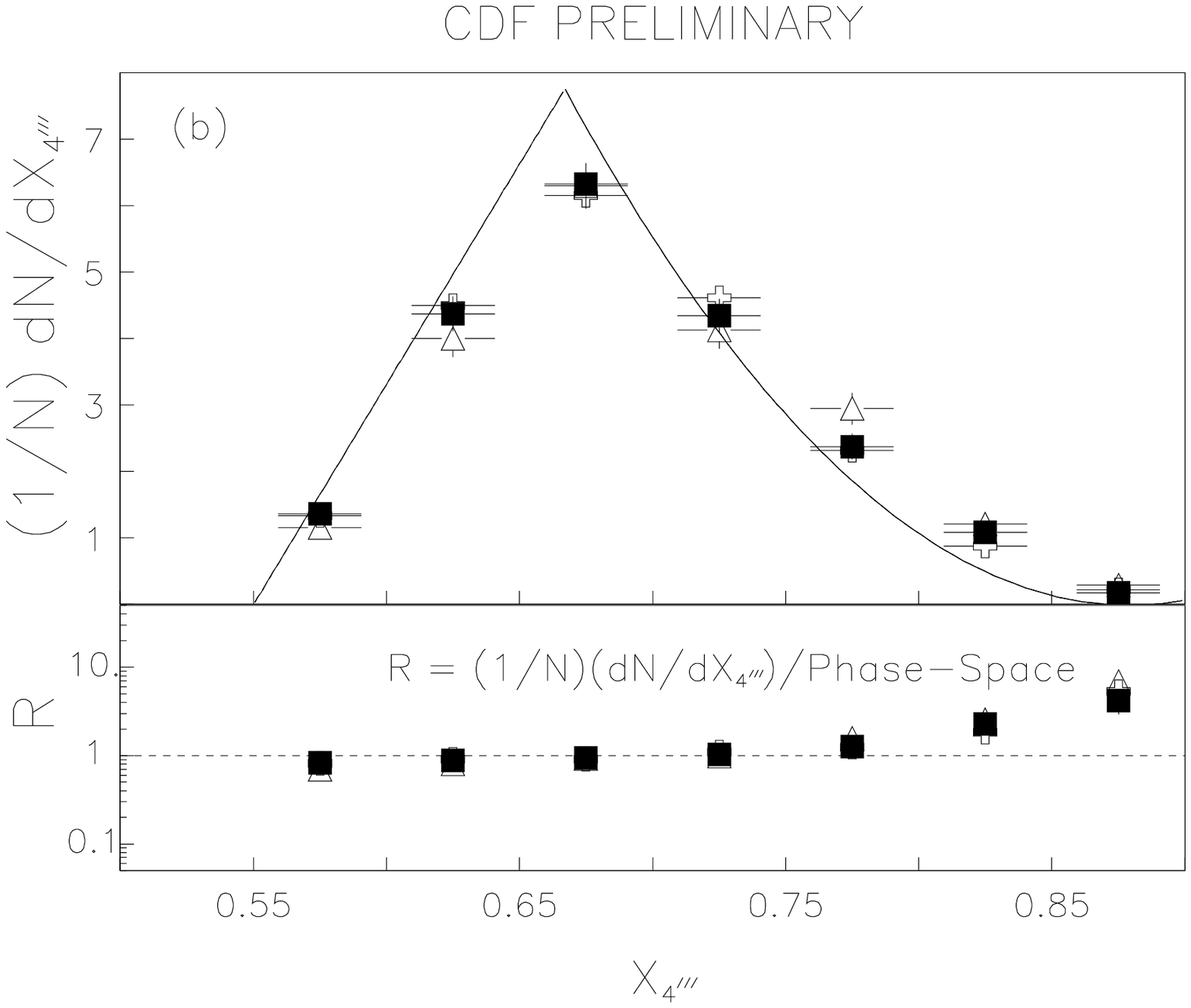,
bb=0 200 550 700,width=8cm,clip=}}
\caption { $X_4$ \em distribution  for the data (solid squares) compared to
           the predictions of HERWIG (triangles), NJETS (crosses) and
           phase space (line). The lower section shows the deviations
           from the prediction of the phase space model. }
\label{figure:x4}
\end{minipage}
\end{figure}

\section{Conclusions}

In conclusion, D0 and CDF have found no evidence, in the datasets 
collected since 1992 at the Tevatron collider, for new physics in multijet
final states. Limits have been put on the mass of all searched
states, except for the Higgs boson, where only cross section limits are yet possible.
The analysis of events with three to six jets in the final state at CDF 
have shown no hints of deviations from the QCD matrix element.  

\end{document}